\documentclass[aps,pre,twocolumn,groupedaddress,superscriptaddress,showpacs]{revtex4-1}

\usepackage{graphicx}
\usepackage{color}
\usepackage{amsmath}
\usepackage{amsfonts}
\usepackage{amssymb}
\usepackage{dcolumn}
\usepackage{hyperref}
\hypersetup{colorlinks=true,urlcolor=blue,linkcolor=blue,citecolor=blue}
\definecolor{orange}{rgb}{1,0.5,0}
\definecolor{green}{rgb}{0,0.5,0}

\begin{document}
\title{Percolation with long-range correlated disorder}
\author{K.~J.~Schrenk}
\email{jschrenk@ethz.ch}
\affiliation{Computational Physics for Engineering Materials, IfB, ETH Zurich, Wolfgang-Pauli-Strasse 27, CH-8093 Zurich, Switzerland}
\author{N.~Pos\'e}
\email{posen@ifb.baug.ethz.ch}
\affiliation{Computational Physics for Engineering Materials, IfB, ETH Zurich, Wolfgang-Pauli-Strasse 27, CH-8093 Zurich, Switzerland}
\author{J.~J.~Kranz}
\email{jkranz@ethz.ch}
\affiliation{Computational Physics for Engineering Materials, IfB, ETH Zurich, Wolfgang-Pauli-Strasse 27, CH-8093 Zurich, Switzerland}
\author{L.~V.~M.~van~Kessenich}
\email{laurensm@ethz.ch}
\affiliation{Computational Physics for Engineering Materials, IfB, ETH Zurich, Wolfgang-Pauli-Strasse 27, CH-8093 Zurich, Switzerland}
\author{N.~A.~M.~Ara\'ujo}
\email{nuno@ethz.ch}
\affiliation{Computational Physics for Engineering Materials, IfB, ETH Zurich, Wolfgang-Pauli-Strasse 27, CH-8093 Zurich, Switzerland}
\author{H.~J.~Herrmann}
\email{hans@ifb.baug.ethz.ch}
\affiliation{Computational Physics for Engineering Materials, IfB, ETH Zurich, Wolfgang-Pauli-Strasse 27, CH-8093 Zurich, Switzerland}
\affiliation{Departamento de F\'{\i}sica, Universidade Federal do Cear\'a, 60451-970 Fortaleza, Cear\'a, Brazil}
\begin{abstract}
Long-range power-law correlated percolation is investigated using Monte Carlo simulations.
We obtain several static and dynamic critical exponents as function of the Hurst exponent $H$ which characterizes the degree of spatial correlation among the occupation of sites.
In particular, we study the fractal dimension of the largest cluster and the scaling behavior of the second moment of the cluster size distribution, as well as the complete and accessible perimeters of the largest cluster.
Concerning the inner structure and transport properties of the largest cluster, we analyze its shortest path, backbone, red sites, and conductivity.
Finally, bridge site growth is also considered.
We propose expressions for the functional dependence of the critical exponents on $H$.
\end{abstract}
\maketitle
%
\section{Introduction}
In percolation on a lattice, each lattice element (site or bond) is occupied with probability $p$ or empty with probability ${1-p}$.
Occupied sites are connected to their nearest neighbors and form clusters, the properties of which depend on $p$ \cite{Stauffer94, Sahimi94}.
There is a threshold value $p_c$, such that for ${p>p_c}$ there exists a cluster spanning between two opposite sides of the lattice.
At ${p=p_c}$, a continuous transition occurs between this connected state and the state for ${p<p_c}$, where there is no spanning cluster.
The spanning cluster is only fractal at ${p=p_c}$.

Percolation theory and related models have been applied to study transport and geometrical properties of disordered systems \cite{Isichenko92, Sahimi93}.
Frequently the disorder in the system under study exhibits power-law long-ranged spatial correlations.
This fact has motivated some studies of percolation models where the sites of the lattice are not occupied independently, but instead with long-range spatial correlation, in a process named \emph{correlated percolation} \cite{Weinrib84, Prakash92, Isichenko92, Sahimi93, Schmittbuhl93, Sahimi94a, Du04, Makse95, Makse98, Sahimi98, Makse00, Araujo02, Araujo03, Sandler04}.
The qualitative picture that emerged from those works is that, in the presence of long-range correlations, percolation clusters become more compact and their transport properties change accordingly.
These findings have also been confirmed by experimental studies of the transport properties of clusters in correlated invasion percolation \cite{Auradou99, Schmittbuhl00}.

The critical exponents of the uncorrelated percolation transition in two dimensions are known rigorously for the triangular lattice \cite{Smirnov01b}.
In addition, at the critical point, the correlation-length diverges and universality holds, i.e., critical exponents and amplitude ratios do not depend on short-range details, such as lattice specifics \cite{Essam80, Isichenko92, Sahimi93, Stauffer94, Sahimi94}.
This statement has been made precise by renormalization group theory, which predicts that the scaling functions within a universality class are the same, while the lattice structure only influences the non-universal metric factors \cite{Privman84, Binney92}.
If, by contrast, infinite-range, power-law correlations are present, according to the extended Harris criterion, the critical exponents can change, depending on how the correlations decay with spatial distance \cite{Weinrib83, Weinrib84, Schmittbuhl93, Sandler04, Janke04b}.

Here, we investigate a two-dimensional percolation model where the sites of a lattice are occupied based on power-law correlated disorder generated with the Fourier filtering method \cite{Peitgen88, Prakash92, Lauritsen93, Makse96, Ballesteros99b, Malamud99, Oliveira11, Ahrens11, Morais11}.
The Hurst exponent $H$ of the disorder is related to the exponent of the power-law decay of spatial correlations with the distance; we find that the fractal dimension of the largest cluster, its perimeter, as well as the dimension of its shortest path, backbone, and red sites depend on $H$
\footnote{
We note that the correlation parameter $\lambda$ in Ref.~\cite{Prakash92} is related to the Hurst exponent $H$ used here by ${\lambda=2(H+1)}$.
For the analogous parameter $a$ of Ref.~\cite{Weinrib84}, one has ${a=-2H}$.
}.
Strong dependence on $H$ is also found for the electrical conductivity exponent of the largest cluster and the growth of bridge sites in the correlated percolation model.
For two-dimensional critical phenomena, conformal field theory has been used to obtain exact values of critical exponents in the form of simple rational numbers \cite{Domb87, DiFrancesco97, Henkel12b}.
Therefore, we make proposals for the functional dependence of all measured exponents on the Hurst exponent $H$, as being the simplest rational expressions that fit the numerical data.

This work is organized as follows.
Section~\ref{sec::correlated_percolation_problem} defines the method of generating long-range correlations and the corresponding correlated percolation model.
In Sec.~\ref{sec::percolation_threshold}, we consider the percolation threshold of the used lattice.
This result is applied in Sec.~\ref{sec::smax_rsm} to measure the fractal dimension of the largest cluster and the scaling behavior of the second moment of the cluster size distribution at the percolation threshold.
The complete and accessible perimeters of the largest cluster are investigated in Sec.~\ref{sec::cluster_perimeters}.
Section~\ref{sec::burning_exponents} discusses shortest path, backbone, and red sites of the largest cluster at the threshold.
The conductivity of the largest cluster is analyzed in Sec.~\ref{sec::conductivity}.
In Sec.~\ref{sec::ranked_surfaces}, we discuss the growth exponent of bridge sites in the correlated percolation model.
Finally, in Sec.~\ref{sec::fin}, we present some concluding remarks.
\section{\label{sec::correlated_percolation_problem}Correlated percolation}
To study correlated percolation on a lattice, it is convenient to work with a landscape of random heights $h$, where $h(\mathbf{x})$ is the height of the landscape at the lattice site at position $\mathbf{x}$ \cite{Weinrib84, Prakash92, Isichenko92, Schmittbuhl93, Kalda08, Kondev00}.
Recently, \emph{ranked surfaces} have been introduced, providing the adequate framework to tackle this problem \cite{Schrenk12}.
The ranked surface of a discrete landscape is constructed as follows:
One first ranks all sites in the landscapes according to their height, from the smallest to the largest value.
Then, a ranked surface is constructed where each site has a number corresponding to its position in the rank.
The following percolation model can then be defined:
Initially, all sites of the ranked surface are unoccupied.
The sites are occupied one-by-one, following the ranking.
At each step, the fraction of occupied sites $p$ increases by the inverse of the total number of sites in the surface.
By this procedure, a configuration of occupied sites is obtained, the properties of which depend on the landscape.
For example, if the heights are distributed uniformly at random, classical percolation with fraction of occupied sites $p$ is obtained \cite{Newman00, Newman01, Hu12}.

Here, we study the case where the heights $h$ have long-range spatial correlations.
Such a power-law correlated disorder can be generated using the Fourier filtering method (Ffm) \cite{Peitgen88, Prakash92, Lauritsen93, Makse96, Ballesteros99b, Malamud99, Sandler04, Oliveira11, Ahrens11, Morais11, Fehr11b}, which is based on the Wiener-Khintchine theorem (WKt) \cite{MacDonald62, Peitgen88}.
The WKt states that the auto-correlation of a time series equals the Fourier transform of its power spectrum, i.e. of the absolute squares of the Fourier coefficients.
This fact is exploited in the Ffm by imposing the following power-law form of the power spectrum $S(\mathbf{f})$ of the disorder:
\begin{equation}
	\label{eqn::power_spectrum}
	S(\mathbf{f}) \sim \vert\mathbf{f}\vert^{-\beta_c} = \left(\sqrt{f_1^2+f_2^2}\right)^{-\beta_c}, 
\end{equation}
where $\beta_c$ defines the Hurst exponent $H$ via ${\beta_c = 2(H+1)}$.
By the WKt, this gives the following correlation function $c(\mathbf{r})$ of the heights $h$,
\begin{equation}
	c(\mathbf{r}) = \left\langle h(\mathbf{x})h(\mathbf{x}+\mathbf{r}) \right\rangle_{\mathbf{x}} \sim \vert\mathbf{r}\vert^{2H}, 
\end{equation}
where the power-law decay of the spatial correlation is described by the Hurst exponent $H$.
For correlated percolation, one considers the range ${-1 \leq H \leq 0}$ \cite{Weinrib84, Isichenko92, Prakash92, Schmittbuhl93}.
${H=-1}$ corresponds to ${\beta_c=0}$, such that the power spectrum in Eq.~(\ref{eqn::power_spectrum}) is independent on the frequency, and the landscape profile is white noise.
This limit recovers uncorrelated percolation.
Since ${H\leq0}$, as $H$ increases towards zero, the correlation function decays more slowly.
In simulations, for a desired value of $H$ one can generate random Fourier coefficients of the heights $h$ with amplitudes according to the power spectrum in Eq.~(\ref{eqn::power_spectrum}) and then apply an inverse fast Fourier transform to obtain $h(\mathbf{x})$ \cite{Peitgen88, Prakash92, Lauritsen93, Makse96, Ballesteros99b, Malamud99, Oliveira11, Ahrens11, Morais11, Fehr11b}.

The extended Harris criterion, as formulated in Refs.~\cite{Weinrib83, Weinrib84, Schmittbuhl93, Sandler04, Janke04b}, states that for the range ${-d/2<H<0}$ the correlations do not affect the critical exponents of the percolation transition if $H \leq -1/\nu_\text{uncorr}$, where $\nu_\text{uncorr}$ is the correlation-length critical exponent and for ${d=2}$, ${\nu_\text{2D}^\text{uncorr}=4/3}$ \cite{Stauffer94, Smirnov01b}.
Whereas for ${-1/\nu_\text{2D}^\text{uncorr}<H<0}$ the critical exponents are expected to depend on the value of $H$.
The quantitative dependence of the critical exponents on $H$, in this regime, is not yet entirely clear.
Concerning the correlation-length critical exponent for the correlated case $\nu_H$, the analytical works in Refs.~\cite{Weinrib83, Weinrib84, Schmittbuhl93} predict that ${\nu_H=-1/H}$.
In the case of Weinrib and Halperin \cite{Weinrib83, Weinrib84} this is a conjecture based on renormalization group calculations; Schmittbuhl \emph{et al.} \cite{Schmittbuhl93} found the same result by analyzing hierarchical networks.
Therefore, in both analytical approaches, it is not certain that $\nu_H$ actually behaves as conjectured and there is some controversy regarding this question, as discussed, e.g., in the field-theoretical work of Prudnikov \emph{et al.} \cite{Prudnikov99, Prudnikov00}.
For correlated percolation, the relation ${\nu_H=-1/H}$ has been supported by the numerical work in Refs.~\cite{Sandler04, Abete04, Marinov06}.
Agreement has also been reported by Prakash \emph{et al.} \cite{Prakash92}, however only approximately for the range ${-1/\nu_\text{2D}^\text{uncorr}\leq H \leq -0.5}$.
Finally, for ${H>0}$ there is no percolation transition \cite{Schmittbuhl93, Olami96}.
In the following, we consider values of the Hurst exponent in the range ${-1 \leq H \leq 0}$.
\section{\label{sec::percolation_threshold}Percolation threshold}
%
\begin{figure}
	\includegraphics[width=\columnwidth]{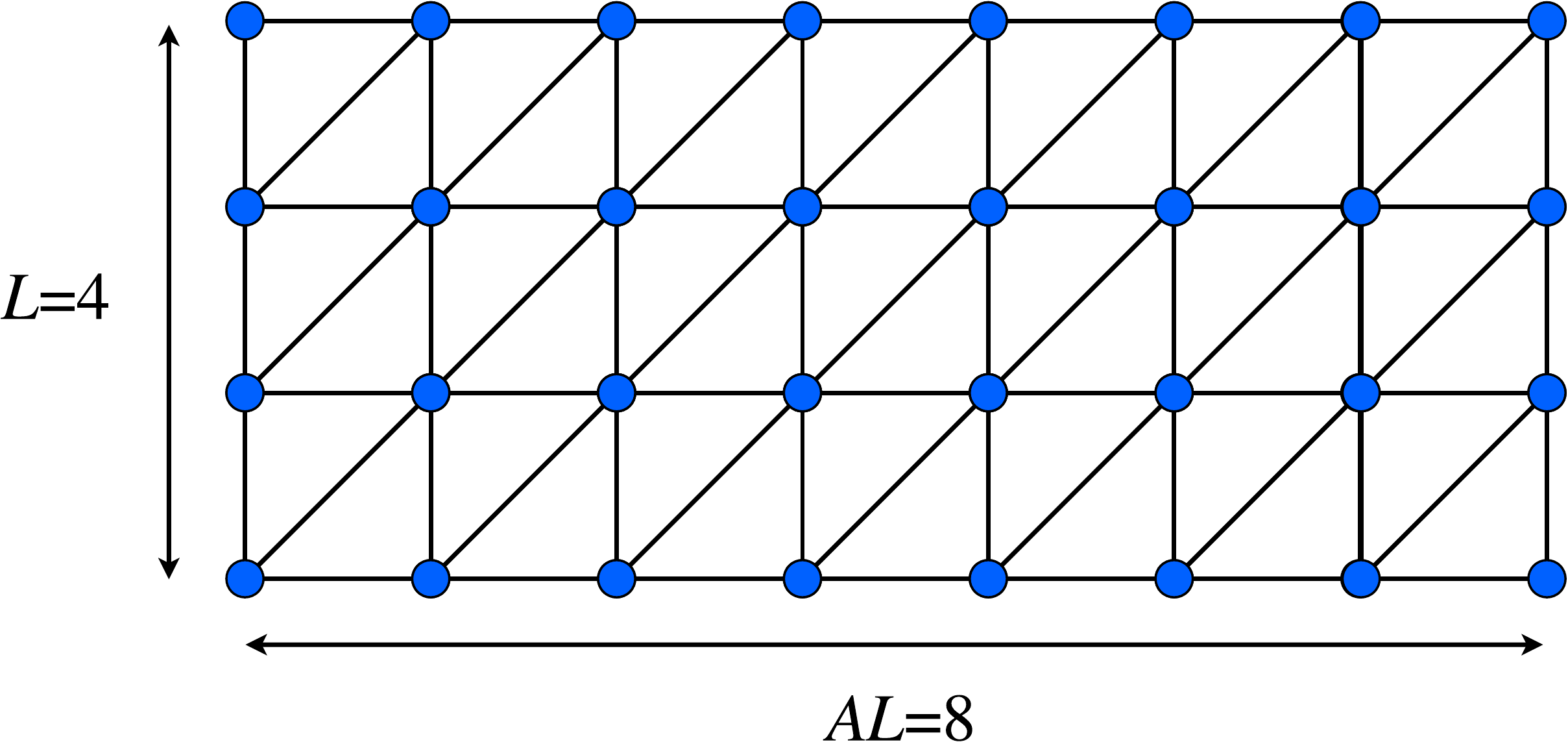}
	\caption{\label{fig::lattice_aspect_definition}
	(Color online)
	Triangular lattice stripe of size ${L=4}$ and aspect ratio ${A=2}$.
	}
\end{figure}
We consider the correlated percolation model defined in Sec.~\ref{sec::correlated_percolation_problem} on triangular lattice stripes of length $L$ and aspect ratio $A$, consisting of ${N=AL^2}$ sites (see Fig.~\ref{fig::lattice_aspect_definition}).
To investigate critical correlated percolation, one first needs to determine the percolation threshold $p_c$ of this lattice.
For site percolation on the triangular lattice, it is possible to show that ${p_c=1/2}$ \cite{Stauffer94}.
The argument of Sykes and Essam \cite{Sykes63, Sykes64b} is as follows:
For certain lattices, one can find their corresponding \emph{matching lattice}.
In the context of Refs.~\cite{Sykes63, Sykes64b}, this is related to matching expansions of the mean number of clusters for high and low $p$.
A more visual explanation of the concept of matching lattice is the following \cite{Isichenko92}:
Suppose that for a lattice $G_1$ there exists a different lattice $G_2$, such that each site in lattice $G_1$ is uniquely related to one site in $G_2$ and the other way around.
Also, assume that if a site is occupied in one of the lattices, its partner in the other one can not be occupied.
Now, if the presence of a cluster spanning $G_2$ in one direction prevents any cluster spanning $G_1$ in the perpendicular direction and, conversely, there can only be a percolating cluster in $G_1$ if there is no percolation in $G_2$, then $G_1$ and $G_2$ are matching lattices.
For example, the triangular lattice is its own matching lattice, called self-matching, while the square lattice is matched by the star lattice \cite{Sykes64b}.
Sykes and Essam argued, based on the uniqueness of the threshold $p_c$ \cite{Sykes63, Sykes64b, Kramers41}, that for any lattice $G_1$ and its matching one $G_2$, the sum of the thresholds of both equals unity:
\begin{equation}
	p_c^{G_1}+p_c^{G_2} = 1.
\end{equation}
Then, since the triangular lattice is self-matching, one has ${p_c^{G_1}=p_c^{G_2}}$ and it follows that ${p_c=1/2}$.
The question of which pairs of lattices match each other is independent on the statistical properties of the heights $h$ that determine the cluster properties.
Therefore, the site percolation threshold of the triangular lattice is ${p_c=1/2}$, also for correlated percolation.
We also checked this statement numerically by measuring $p_c$ for different values of the Hurst exponent $H$, finding that it is compatible with ${1/2}$, within error bars
\footnote{
We also determined ${p_c^\text{square}}$ of the square and ${p_c^\text{star}}$ of the star lattice for various values of $H$ and found that, in contrast to the behavior of the triangular lattice, the percolation threshold does depend on $H$.
Our results for ${p_c^\text{square}(H)}$ of the square lattice agree, within error bars, with the ones reported in Ref.~\cite{Prakash92}.
Also, we have, within error bars, ${ p_c^\text{square}(H) + p_c^\text{star}(H) = 1 }$, consistent with the matching property.
Besides ${p_c=1/2}$, an additional advantage of the triangular lattice is that the cluster perimeters (see Sec.~\ref{sec::cluster_perimeters}) are well-defined, avoiding common problems encountered on the square lattice \cite{Grossman87, Saberi09b, Adams10, Chatelain10}.
}.

\begin{figure}
	\includegraphics[width=\columnwidth]{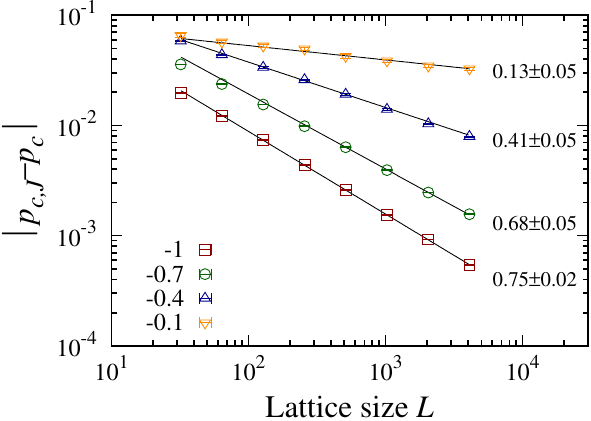}\\
	\caption{
	\label{fig::other_estimators_pcj}
	(Color online)
	Convergence of the percolation threshold estimator $p_{c,J}$.
	The difference between the estimator and the threshold ${\vert p_{c,J} - 1/2 \vert}$ is shown as function of the lattice size $L$ for $H=-1$, $-0.7$, $-0.4$, and $-0.1$.
	The data is shifted vertically to improve visibility.
	Results are averages over $10^5$ samples.
	We keep track of the cluster properties with the labeling method proposed by Newman and Ziff \cite{Newman00, Newman01}, as in Ref.~\cite{Hoshen76}.
 	}
\end{figure}

As a first check of the theory presented in Refs.~\cite{Weinrib83, Weinrib84, Schmittbuhl93} regarding the dependence of $\nu_H$ on $H$, we consider here the convergence of a threshold estimator, namely the value ${p_{c,J}}$ at which the maximum change in the size of the largest cluster $s_\text{max}$ occurs \cite{Nagler11, Chen11, Manna11, Schrenk11b, Reis11, Zhang12b, Cho13, Chen13}.
The expected scaling behavior \cite{Manna11, Schrenk13} is
\begin{equation}
	\vert p_{c,J}(L) - p_c \vert \sim L^{-1/\nu_H}, 
\end{equation}
where ${p_c=1/2}$.
Figure \ref{fig::other_estimators_pcj} shows ${\vert p_{c,J}(L) - p_c \vert}$ as function of the lattice size $L$ for different values of $H$.
Within error bars, the data is compatible with ${1/\nu_H=-H}$ for the considered values of $H$.
%
\section{\label{sec::smax_rsm}Maximum cluster size and second moment}
\begin{figure}
	\begin{tabular}{l}
		(a)\\
		\includegraphics[width=\columnwidth]{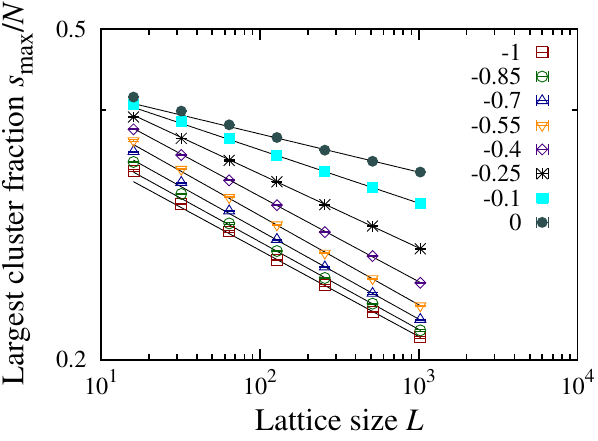}\\
		(b)\\
		\includegraphics[width=\columnwidth]{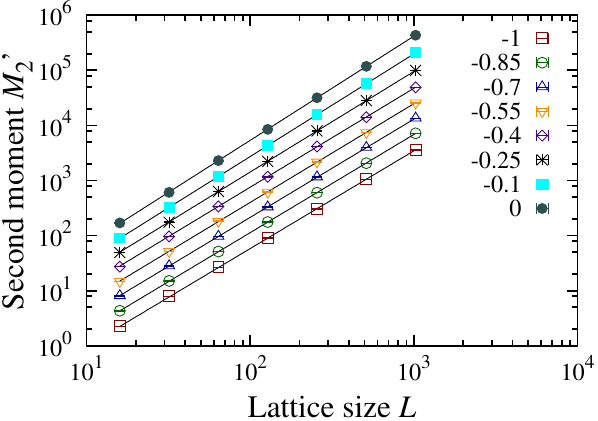}\\
	\end{tabular}
	\caption{\label{fig::smax_sec_mom_log_log}(Color online)
	(a) Fraction of sites in the largest cluster ${s_\text{max}/N}$ as function of the lattice size $L$ for different values of $H$.
	(b) Second moment of the cluster size distribution $M_2'$ as function of $L$ for the same values of $H$ as in (a).
	The data is shifted vertically to improve visibility.
	Solid black lines are guides to the eye.
	Results are averages over at least $10^4$ samples.
	}
\end{figure}

\begin{figure}
	\begin{tabular}{l}
	(a)\\
	\includegraphics[width=\columnwidth]{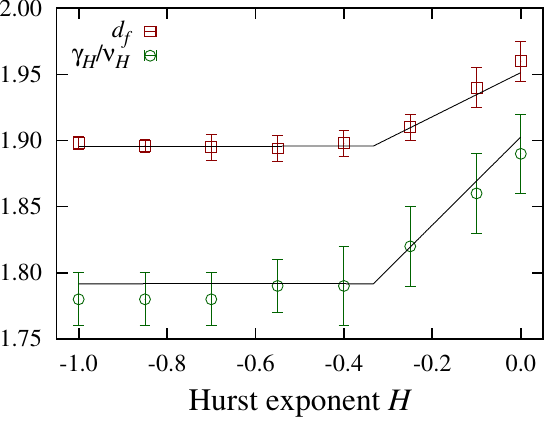}\\
	(b)\\
	\includegraphics[width=\columnwidth]{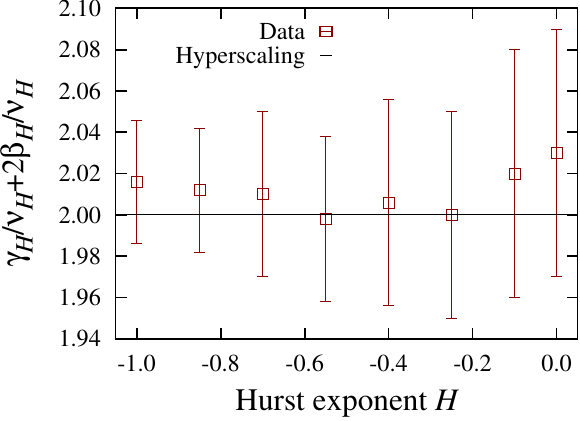}\\
	\end{tabular}
	\caption{\label{fig::df_gamma_nu_hyperscaling}
	(Color online)
	(a) Fractal dimension of the largest cluster $d_f$ and critical exponent ratio ${\gamma_H/\nu_H}$ as function of the Hurst exponent $H$, where $\gamma_H$ is the susceptibility critical exponent and $\nu_H$ is the correlation-length critical exponent.
	For ${H>-1/3}$, the solid lines show the expressions of Eqs.~(\ref{eqn::local_slopes_df_L}) and (\ref{eqn::conj_gamma_nu}).
	(b) With ${d_f=d-\beta_H/\nu_H}$, where $\beta_H$ is the order parameter critical exponent, the hyperscaling relation reads: ${2=d=\gamma_H/\nu_H+2\beta_H/\nu_H}$ \cite{Stauffer94}.
	One observes that the data agrees, within error bars, with the hyperscaling relation.
	}
\end{figure}

At the threshold, ${p=p_c}$, the largest cluster is a fractal of fractal dimension $d_f$, i.e., its size $s_\text{max}$ scales with the lattice size $L$ as
\begin{equation}
	\label{eqn::smax_scaling_L}
	s_\text{max} \sim L^{d_f}.
\end{equation}
This is also related to the order parameter $P_\infty$ of the percolation transition, which is defined as the fraction of sites in the largest cluster,
\begin{equation}
	P_\infty = s_\text{max}/N,
\end{equation}
and is expected to scale at ${p=p_c}$ as
\begin{equation}
	P_\infty \sim L^{-\beta/\nu} = L^{d_f-d},
\end{equation}
where $\beta$ is the order parameter critical exponent and ${d=2}$ is the spatial dimension \cite{Stauffer94}.
For uncorrelated percolation, ${\beta=5/36}$ and ${\nu=\nu_\text{2D}^\text{uncorr}=4/3}$, such that ${d_f=91/48\approx1.8958}$ \cite{Stauffer94}.
To measure $d_f$ as function of $H$, we considered the scaling of the size of the largest cluster $s_\text{max}$ with the lattice size [see Fig.~\ref{fig::smax_sec_mom_log_log}(a) and Eq.~(\ref{eqn::smax_scaling_L})].
For different values of $H$, we measured $s_\text{max}(L)$ and calculated the local slopes $d_f(L)$ of the data (see e.g. Ref.~\cite{Grassberger99}),
\begin{equation}
	\label{eqn::local_slopes_df_L}
	d_f(L) = \log[ s_\text{max}(2L)/s_\text{max}(L/2) ]/\log(4).
\end{equation}
Finally, $d_f(L)$ is extrapolated to the thermodynamic limit, ${L\to\infty}$, to obtain $d_f(H)$, see Fig.~\ref{fig::df_gamma_nu_hyperscaling}(a).
The fractal dimension is, within error bars, independent on $H$, for ${H\lesssim-1/3}$.
For $H$ approaching zero, the value of $d_f$ does increase.
While this behavior is in agreement with Ref.~\cite{Prakash92}, it is in strong contrast to the behavior of all other fractal dimensions considered in this work, whose values depend strongly on $H$.
Based on the data, we propose the following dependence of $d_f$ on $H$ (in the range ${-1/3 \leq H \leq 0}$) as being the simplest rational expression that fits the numerical data:
\begin{equation}
	\label{eqn::conj_df_smax}
	d_f(H) = \frac{91}{48}+\frac{13}{80}\left( \frac{1}{3}+H \right).
\end{equation}

The hyperscaling,
\begin{equation}
	\label{eqn::hyperscaling}
	d = \frac{\gamma}{\nu}+2\frac{\beta}{\nu}
	= \frac{\gamma}{\nu}+2(d-d_f),
\end{equation}
relates the fractal dimension $d_f$ to the susceptibility critical exponent $\gamma$ and the correlation-length critical exponent $\nu$ \cite{Stauffer94}.
For uncorrelated percolation, ${\gamma=43/18}$ and therefore ${\gamma/\nu_\text{2D}^\text{uncorr}=43/24\approx1.7917}$ \cite{Stauffer94}.
To test the validity of Eq.~(\ref{eqn::hyperscaling}) for different values of $H$, we measure ${\gamma_H/\nu_H}$, where $\gamma_H$ and $\nu_H$ are the susceptibility and correlation length critical exponents for a certain $H$, by considering the scaling behavior of the second moment $M_2'$, defined as,
\begin{equation}
	\label{eqn::rsm_definition}
	M_2' = M_2 - s_\text{max}^2/N,
\end{equation}
where
\begin{equation}
	M_2 = \sum_{k} s_k^2/N, 
\end{equation}
and the sum goes over all clusters with $s_k$ being the number of sites in cluster $k$.
At ${p=p_c}$, the following scaling with the lattice size $L$ is expected \cite{Stauffer94, Ziff10}:
\begin{equation}
	M_2' \sim L^{\gamma_H/\nu_H}.
\end{equation}
In Fig.~\ref{fig::smax_sec_mom_log_log}(b), one sees $M_2'$ as function of $L$ for different values of $H$.
Figure \ref{fig::df_gamma_nu_hyperscaling}(a) shows ${\gamma_H/\nu_H}$, while $\gamma_H/\nu_H+2\beta_H/\nu_H$ is plotted in Fig.~\ref{fig::df_gamma_nu_hyperscaling}(b) for different values of $H$.
One observes that the hyperscaling relation, Eq.~(\ref{eqn::hyperscaling}), is fulfilled, within error bars.
Based on this result, we propose for the functional dependence of $\gamma_H/\nu_H$ on the Hurst exponent $H$, in the range ${-1/3 \leq H \leq 0}$, as being the simplest rational expression that fits the numerical data:
\begin{equation}
	\label{eqn::conj_gamma_nu}
	\frac{\gamma_H}{\nu_H} = (76+13H)/40.
\end{equation}
\section{\label{sec::cluster_perimeters}Cluster perimeters}
\begin{figure}
	\includegraphics[width=\columnwidth]{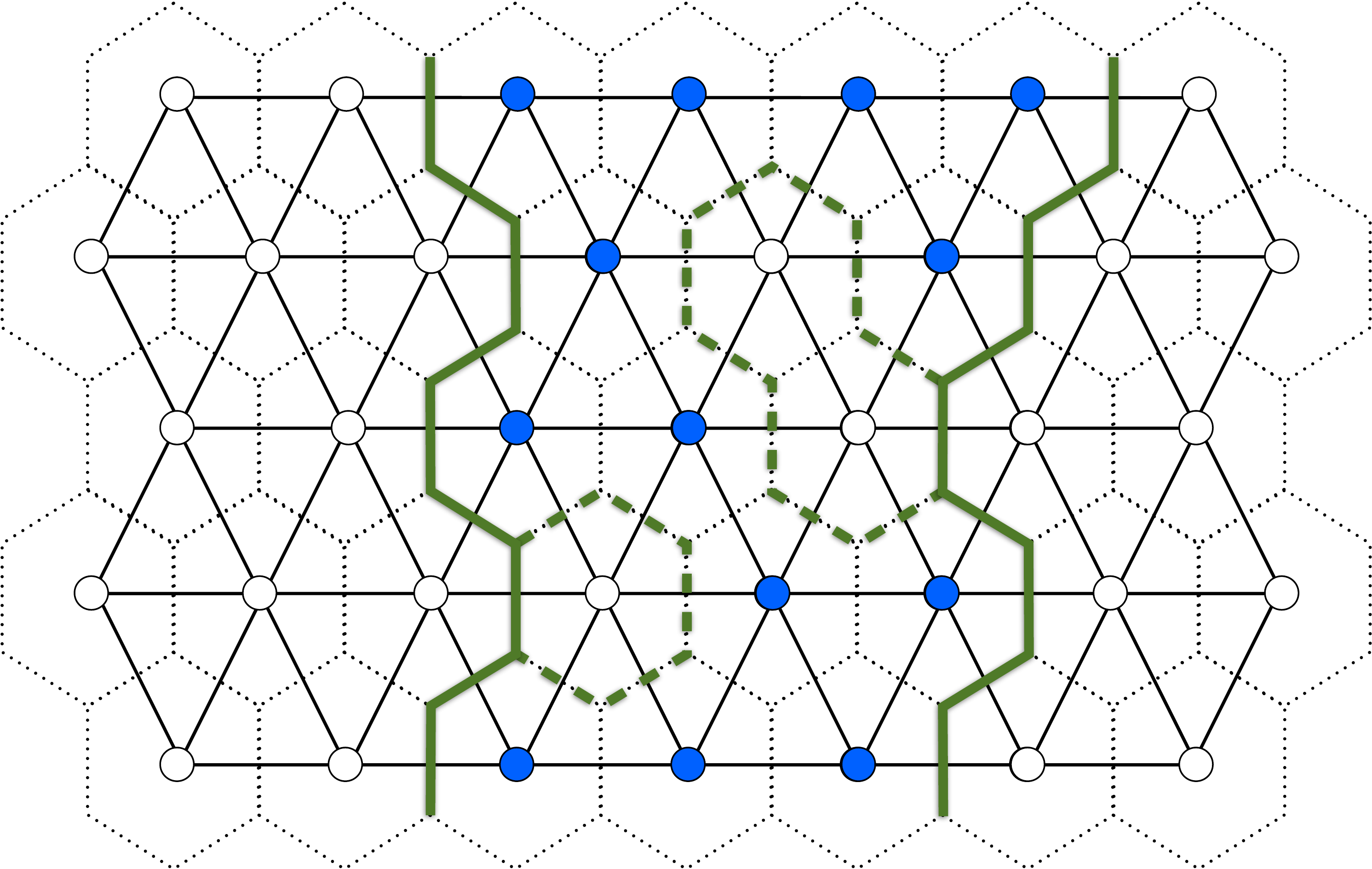}
	\caption{\label{fig::perimeters_honeycomb_defs}
	(Color online)
	Complete and accessible perimeter.
	The blue (filled) sites of the triangular lattice are part of the largest cluster, while the white (empty) sites are unoccupied.
	Bonds of the dual lattice are shown as dashed lines.
	Assume that the largest cluster percolates in the vertical direction and does not touch the left or right boundaries of the lattice.
	Consider a walker starting on the left-bottom side of the lattice, which never visits a bond twice and traces out the complete perimeter, turning left or right depending on which of the two available bonds separates an occupied from an empty site.
	The complete perimeter is fully determined when the top side of the lattice is reached.
	Performing the same walk, but with the additional constraint that fjords with diameter ${\leq\sqrt{3}/3}$ (in lattice units) are not accessible, yields the accessible perimeter.
	The solid green (thick) lines on the honeycomb lattice form the accessible perimeter, while dashed green (thick) lines indicate bonds that are part of the complete perimeter but not of the accessible one.
	A similar walk yields the two perimeters on the right hand side of the cluster.
	}
\end{figure}

\begin{figure*}
	\begin{tabular}{ll}
	(a) ${H=-1}$ &
	(b) ${H=-0.5}$ \\
	\includegraphics[width=\columnwidth]{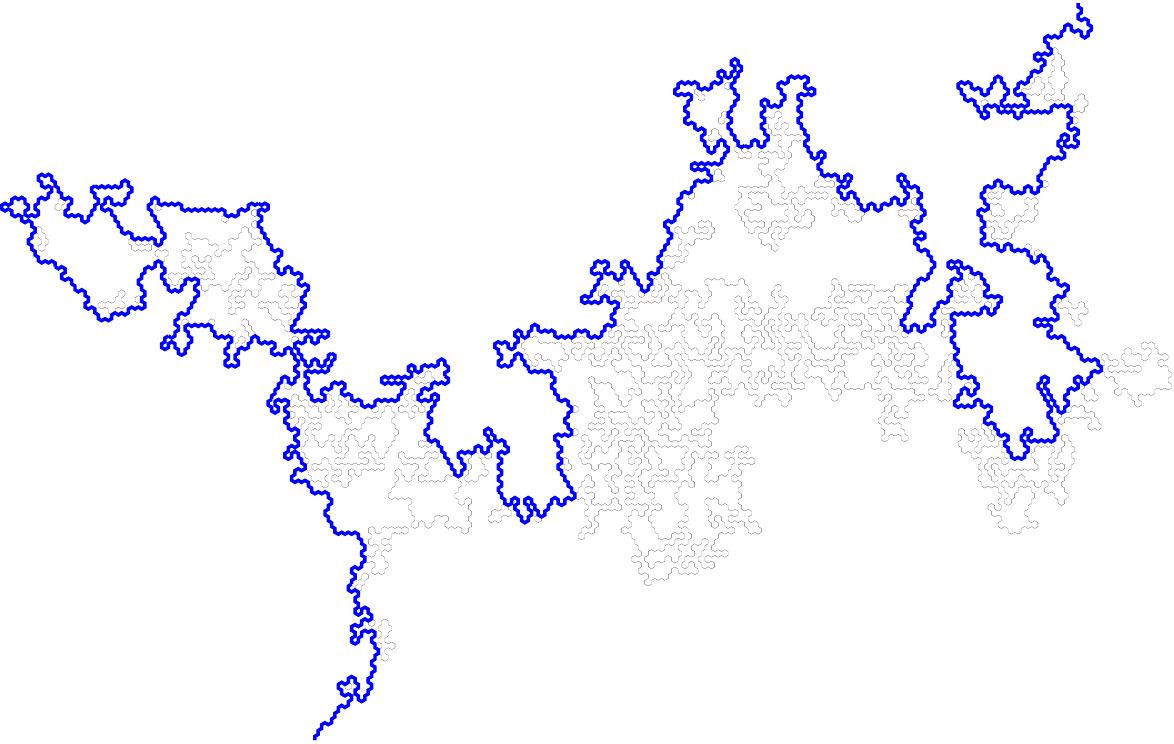} &
	\includegraphics[width=\columnwidth]{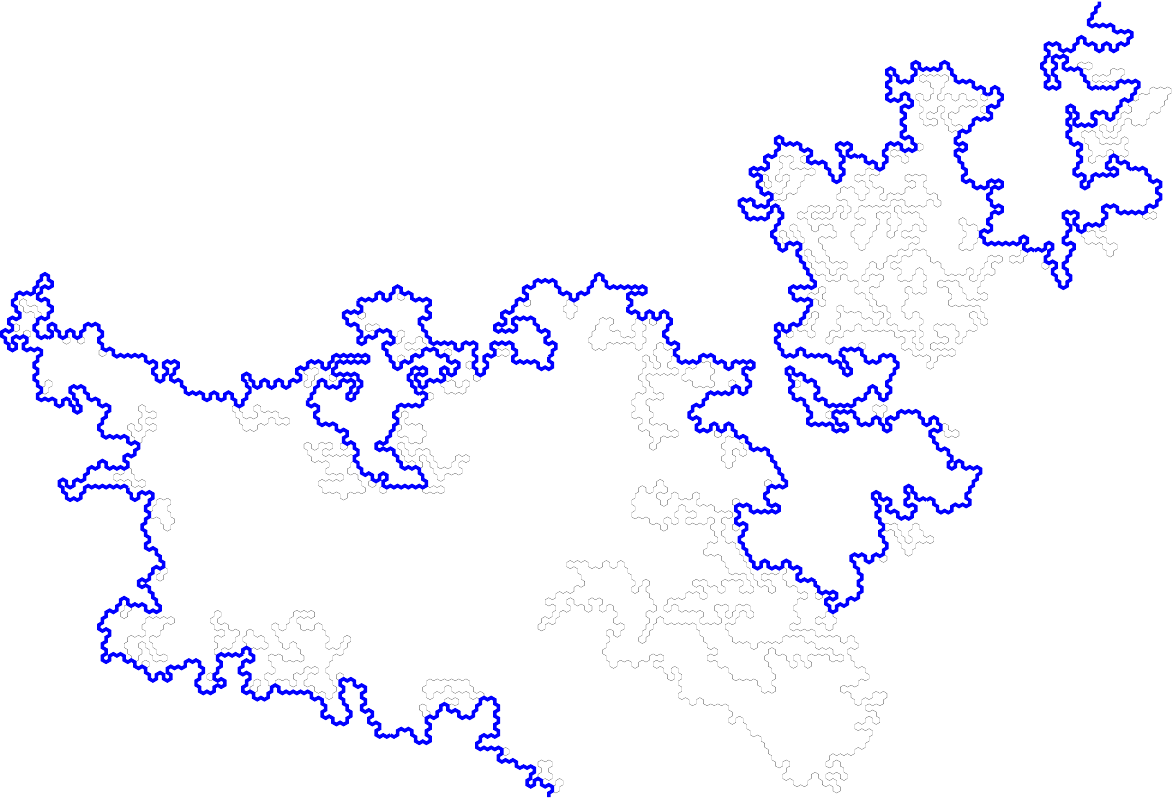} \\
	(c) ${H=-0.25}$ &
	(d) ${H=0}$ \\
	\includegraphics[width=\columnwidth]{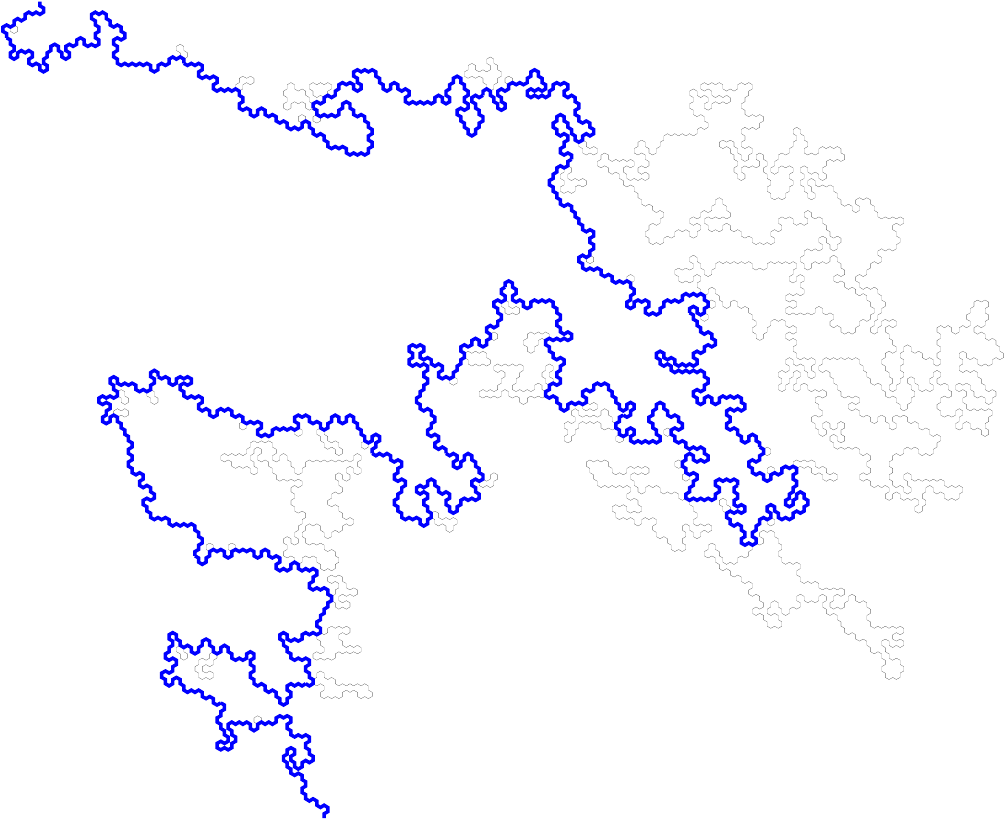} & 
	\includegraphics[width=\columnwidth]{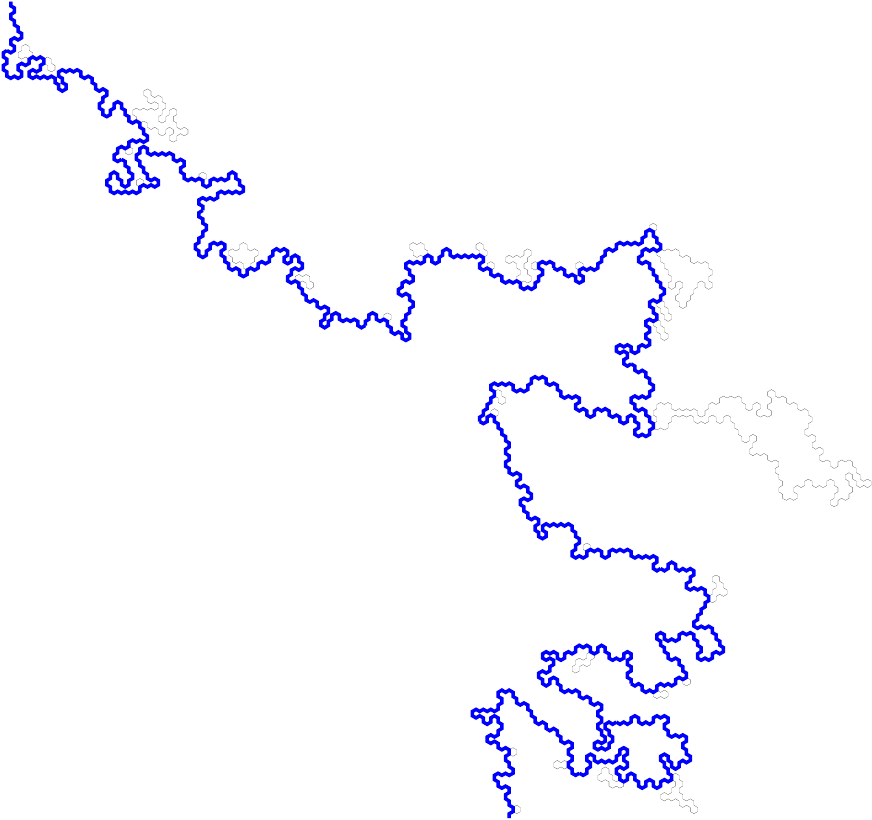} \\
	\end{tabular}
	\caption{\label{fig::perimeter_snaps}
	(Color online)
	Snapshots of typical complete and accessible perimeters.
	The accessible perimeter is shown in bold solid blue lines.
	In addition, the parts of the complete perimeter that do not belong to the accessible perimeter are drawn with thin black lines.
	The snapshots are taken for (a) ${H=-1}$, (b) $-0.5$, (c) $-0.25$, and (d) $0$, on a lattice of (vertical) length ${L=128}$.
	}
\end{figure*}

\begin{figure}
	\includegraphics[width=\columnwidth]{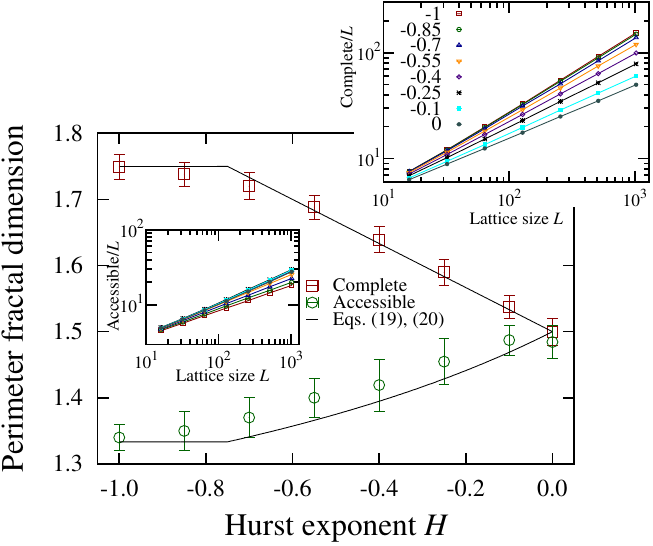}\\
	\caption{\label{fig::perimeter_exponents_hurst}
	(Color online)
	Main plot: Fractal dimension of the complete perimeter $d_\text{cp}$ and of the accessible perimeter $d_\text{ap}$ as function of the Hurst exponent $H$.
	For ${H=-1}$ (uncorrelated), our results, ${d_\text{cp}=1.75\pm0.02}$ and ${d_\text{ap}=1.34\pm0.02}$, are in agreement with values previously reported \cite{Voss84, Sapoval84, Grassberger86b, Grossman87, Saleur87, Aizenman99, Duplantier00}.
	With increasing $H$, both fractal dimensions seem to approach ${3/2}$, compatible with the data of Kalda \emph{et al.} \cite{Kondev95, Kalda01, Kalda08, Mandre11}.
	In the range ${-1/\nu_\text{2D}^\text{uncorr} \leq H \leq 0}$, the solid lines show the expressions
	$d_\text{cp} = {3}/{2}-H/3$
	and
	${d_\text{ap} = {(9-4H)}/{(6-4H)}}$.
	Insets: Length of the complete and of the accessible perimeter as function of the lattice size $L$ for the values of $H$ shown in the main plot.
	}
\end{figure}

\begin{figure}
	\includegraphics[width=\columnwidth]{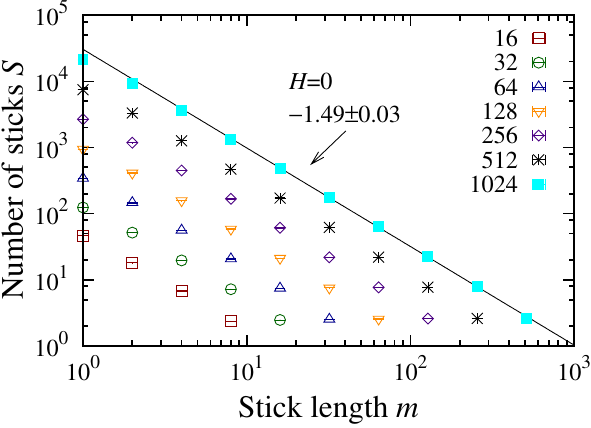}
	\caption{\label{fig::complete_perimeter_yardsticks}
	(Color online)
	Yardstick method to measure the fractal dimension of the complete perimeter.
	The number of sticks needed to follow the perimeter $S$ is shown as function of the stick length $m$, for different lattice sizes $L$, and ${H=0}$.
	The numerical value of the complete perimeter fractal dimension ${d_\text{cp}(H)}$ obtained with the yardstick method,
	${d_\text{cp}(0)=1.49\pm0.03}$,
	agrees, within error bars, with the results of the analysis of the local slopes of the perimeter length (see Fig.~\ref{fig::perimeter_exponents_hurst}), as well as with the literature \cite{Kalda01, Kalda08, Mandre11}.
	}
\end{figure}

Here, we consider triangular lattice stripes of aspect ratio ${A=8}$ (Fig.~\ref{fig::lattice_aspect_definition}).
For every largest cluster that spans the lattice vertically (between the long sides of the lattice, Fig.~\ref{fig::lattice_aspect_definition}) and does not touch its vertical boundaries, there are two contours that can be defined: the complete and accessible perimeters \cite{Voss84, Sapoval84, Ziff84, Weinrib85, Grassberger86b, Lawler01, Lawler01b, Asikainen03, Adams10, Kalda01, Kalda08, Mandre11}.
Figure \ref{fig::perimeters_honeycomb_defs} shows the definition of the two perimeters, which live on the honeycomb lattice, in the case of the triangular lattice:
The complete perimeter consists of all bonds of the honeycomb lattice that separate sites belonging to the spanning cluster from unoccupied sites that can be reached from the vertical boundaries of the lattice without crossing sites belonging to the largest cluster.
If, in addition, fjords of the perimeter with diameter less than $\sqrt{3}/3$ (lattice units) are inaccessible, the accessible perimeter is obtained.
Figure \ref{fig::perimeter_snaps} shows the left hand side complete and accessible perimeters of a percolating cluster on a lattice of size ${L=128}$.
In the upper inset of Fig.~\ref{fig::perimeter_exponents_hurst}, the length of the complete perimeter $M_\text{cp}$ is observed to scale with the lattice size $L$ as,
\begin{equation}
	\label{eqn::complete_perimeter_mass_scaling}
	M_\text{cp} \sim L^{d_\text{cp}},
\end{equation}
where for the uncorrelated case, given by ${H=-1}$, it is known that ${d_\text{cp}=7/4}$ \cite{Voss84, Sapoval84, Lawler01, Lawler01b}.
In addition to considering the scaling of $M_\text{cp}$ with $L$, we also determined the fractal dimension $d_\text{cp}$ using the yardstick method \cite{Mandelbrot83, Tricot88}.
There, one measures the number of sticks ${S(m)}$ of size $m$ needed to follow the perimeter from one end to the other.
Figure \ref{fig::complete_perimeter_yardsticks} shows that, for intermediate stick lengths, ${S(m)}$ scales as
\begin{equation}
	S \sim m^{-d_\text{cp}}.
\end{equation}
We measured the value of the fractal dimension with this method for different lattice sizes $L$ (see Fig.~\ref{fig::complete_perimeter_yardsticks}) and then extrapolated the results to ${L\to\infty}$ to obtain $d_\text{cp}$.
The fractal dimension ${d_\text{cp}(H)}$ determined by this method is compatible with the one obtained from the scaling of the length of the perimeter [see Eq.~(\ref{eqn::complete_perimeter_mass_scaling})] and we combined both measurements for the final estimates:
In Fig.~\ref{fig::perimeter_exponents_hurst}, one sees the fractal dimension of the complete perimeter as function of the $H$.
For $H$ approaching zero, $d_\text{cp}$ decreases and finally converges towards ${3/2}$, in agreement with previous results \cite{Kalda01, Kalda08, Mandre11}.

The fractal dimension of the accessible perimeter $d_\text{ap}$ is defined by the scaling of the length of the accessible perimeter $M_\text{ap}$ with $L$ (see lower inset of Fig.~\ref{fig::perimeter_exponents_hurst}),
\begin{equation}
	M_\text{ap} \sim L^{d_\text{ap}}.
\end{equation}
For uncorrelated percolation the fractal dimension of the accessible perimeter is known to be ${d_\text{ap}=4/3}$ \cite{Grossman87, Saleur87, Lawler01, Lawler01b}.
Figure \ref{fig::perimeter_exponents_hurst} shows $d_\text{ap}(H)$, determined using the scaling of $M_\text{ap}$ and the yardstick method.

\begin{figure}
	\includegraphics[width=\columnwidth]{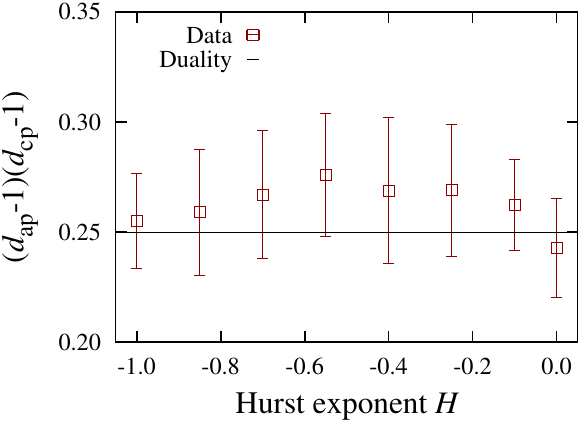}
	\caption{\label{fig::duplantier_duality}
	(Color online)
	Left hand side of the duality relation for cluster perimeters, ${(d_\text{ap}-1)(d_\text{cp}-1)=1/4}$ \cite{Duplantier00, Janke04}, as function of the Hurst exponent $H$.
	}
\end{figure}

For the critical $Q$-state Potts model \cite{Wu82}, Duplantier \cite{Duplantier00, Janke04} established the following duality relation between the fractal dimension of the complete perimeter $d_\text{cp}$ and of the accessible perimeter $d_\text{ap}$:
\begin{equation}
	\label{eqn::duplantier_duality}
	(d_\text{ap}-1)(d_\text{cp}-1)=1/4.
\end{equation}
The case ${Q=1}$ corresponds to uncorrelated percolation \cite{Fortuin72}.
Having measured $d_\text{cp}$ and $d_\text{ap}$ as functions of $H$, we see in Fig.~\ref{fig::duplantier_duality} that the duality relation of Eq.~(\ref{eqn::duplantier_duality}) holds, within error bars, for ${-1 \leq H \leq 0}$.
Therefore, taking the known results for ${H=-1}$ and ${H=0}$ into account, we propose the following functional dependence of the complete perimeter fractal dimension on $H$ (in the range ${-1/\nu_\text{2D}^\text{uncorr}\leq H \leq 0}$, see Ref.~\cite{Mandre11}):
\begin{equation}
	d_\text{cp} = \frac{3}{2}-\frac{H}{3},
\end{equation}
which, assuming the validity of the duality relation also for correlated percolation, implies the following form of the accessible perimeter fractal dimension:
\begin{equation}
	d_\text{ap} = \frac{9-4H}{6-4H}.
\end{equation}
%
\section{\label{sec::burning_exponents}Shortest path, backbone, and red sites}
\begin{figure}
	\includegraphics[width=\columnwidth]{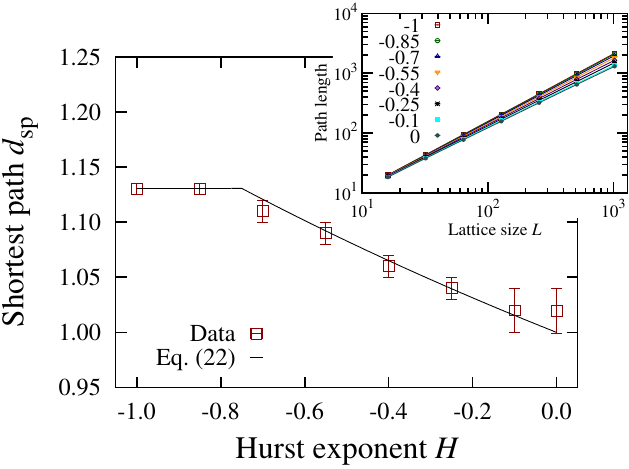}
	\caption{
	\label{fig::03_shortest_path_length}
	(Color online)
	Fractal dimension of the shortest path $d_\text{sp}$ of the largest cluster as function of the Hurst exponent $H$.
	The inset shows the number of sites in the shortest path as function of the lattice size $L$ for the same value of $H$ as in the main plot.
	For uncorrelated disorder, i.e. ${H=-1}$, we find ${d_\text{sp}=1.130\pm0.005}$, in agreement with the literature \cite{Herrmann84, Herrmann88, Grassberger92b, Zhou12}.
	With increasing Hurst exponent, $d_\text{sp}$ approaches unity \cite{Makse96b}.
	This behavior is due to the backbone becoming increasingly compact as $H$ approaches $0$, see Fig.~\ref{fig::06_backbone_size}.
	The solid line is the graph of the proposed behavior of the shortest path fractal dimension: 
	${d_\text{sp}(H) = {147}/{130}-{(3/4+H)}/{(195/34+H)}}$, for ${-3/4 \leq H \leq 0}$,
	and
	$d_\text{sp}(-1\leq H \leq -1/\nu_\text{2D}^\text{uncorr}) = d_\text{sp}(-1/\nu_\text{2D}^\text{uncorr})$.
	}
\end{figure}

For uncorrelated percolation, the shortest path between two sites in the largest cluster is a fractal of dimension ${d_\text{sp}\approx1.131}$ \cite{Herrmann84, Herrmann88, Grassberger92b, Zhou12}.
For a given configuration, it can be identified using the burning method \cite{Herrmann84}:
On the cluster spanning the lattice vertically [with aspect ratio ${A=1}$ (see Fig.~\ref{fig::lattice_aspect_definition})], we select one cluster site in the top row and one in the bottom row, such that their Euclidean distance is minimized and find the number of sites $M_\text{sp}$ in the shortest path between them.
The following scaling of the length with the lattice size $L$ is observed:
\begin{equation}
	M_\text{sp} \sim L^{d_\text{sp}},
\end{equation}
which can be used to determine the fractal dimension ${d_\text{sp}(H)}$ using the local slopes [see Eq.~(\ref{eqn::local_slopes_df_L})], shown in Fig.~\ref{fig::03_shortest_path_length}.
These results are also compatible with the ones obtained using the yardstick method (not shown).
For increasing correlation, $d_\text{sp}$ deceases and is compatible with unity for ${H=0}$, as also reported in Ref.~\cite{Makse96b}.
Using this observation and the literature results for uncorrelated percolation \cite{Herrmann84, Herrmann88, Grassberger92b, Zhou12}, we propose the following dependence of $d_\text{sp}$ on the Hurst exponent $H$ (in the range ${-1/\nu_\text{2D}^\text{uncorr}\leq H \leq 0}$):
\begin{equation}
	d_\text{sp}(H) = \frac{147}{130}-\frac{3/4+H}{195/34+H}.
\end{equation}

\begin{figure}
	\includegraphics[width=\columnwidth]{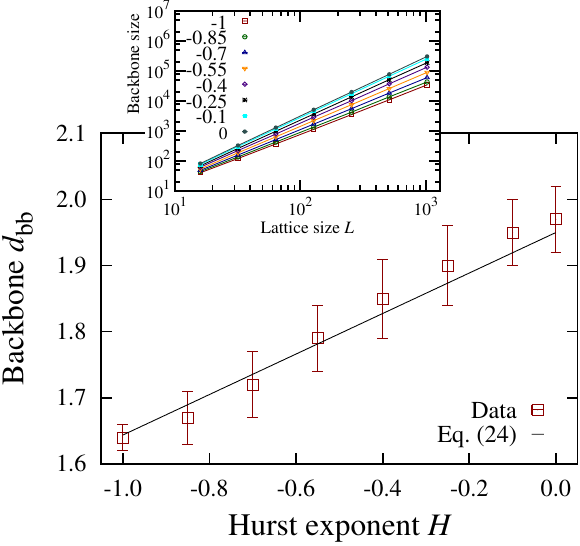}
	\caption{
	\label{fig::06_backbone_size}
	(Color online)
	Fractal dimension of the backbone $d_\text{bb}$ as function of the Hurst exponent $H$.
	With increasing $H$, the backbone becomes more compact and, consequently, $d_\text{bb}$ increases, while the fractal dimension of the shortest path (see Fig.~\ref{fig::03_shortest_path_length}) decreases \cite{Prakash92}.
	For uncorrelated disorder, ${H=-1}$, we measure ${d_\text{bb}=1.64\pm0.02}$, compatible with the results reported in Refs.~\cite{Herrmann84b, Herrmann84, Grassberger92b, Rintoul92, Grassberger99, Deng04c}.
	The solid line is the graph of the following interpolation:
	$d_\text{bb}(H)={39}/{20}(1+H)-{166}/{101}H$.
	Inset: Backbone size as function of the lattice size $L$ for the same values of $H$ as in the main plot.
	}
\end{figure}

In addition to measuring the length of the shortest path between two sites in the largest cluster, one can also ask which sites would carry non-zero current if the occupied sites would be resistors and a potential difference were applied between these two sites.
This subset of sites of the largest cluster is called backbone and it is the union of all non-self-crossing paths between these two sites \cite{Stauffer94, Herrmann84b, Herrmann84, Grassberger92b, Rintoul92, Grassberger99, Lawler01c, Deng04c}.
Some sites of the backbone are singly connected, i.e., the connectivity between the two ends of the backbone is broken if any one of these sites is removed.
These sites are called red sites \cite{Coniglio89, Scholder09, Schrenk12}.
Algorithmically, for a given cluster, the backbone and its red sites can be found with the burning method \cite{Herrmann84}.
The total number of sites in the backbone $M_\text{bb}$ scales with the lattice size $L$,
\begin{equation}
	M_\text{bb} \sim L^{d_\text{bb}},
\end{equation}
where $d_\text{bb}$ is the backbone fractal dimension, see inset of Fig.~\ref{fig::06_backbone_size}.
With increasing $H$, $d_\text{bb}$ increases and is compatible with the fractal dimension of the largest cluster for $H$ approaching zero.
Similarly to Ref.~\cite{Prakash92}, for the functional dependence of $d_\text{bb}$ on $H$, we propose to interpolate linearly between the best known value for uncorrelated percolation, ${d_\text{bb}(-1)=1.6434\pm0.0002}$ \cite{Deng04c} and the fractal dimension of the largest cluster for ${H=0}$ [see Eq.~(\ref{eqn::conj_df_smax})]:
\begin{equation}
	d_\text{bb}(H)=\frac{39}{20}(1+H)-\frac{166}{101}H.
\end{equation}

\begin{figure}
	\includegraphics[width=\columnwidth]{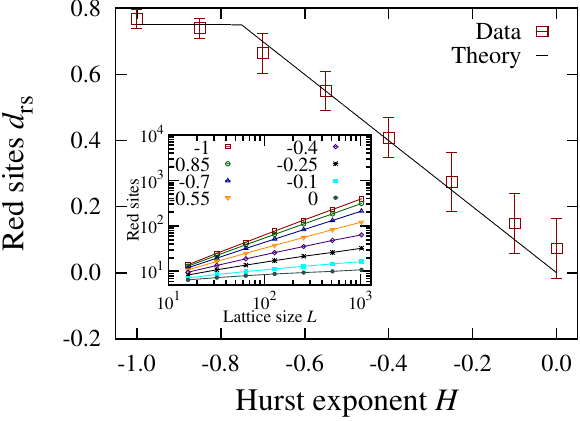}
	\caption{\label{fig::red_sites_backbone}
	(Color online)
	Fractal dimension of the red sites $d_\text{rs}$ as function of the Hurst exponent $H$.
	In the main plot, based on the result by Coniglio \cite{Coniglio89, Scholder09, Schrenk12}, the data (squares) is compared to the theoretical prediction for ${1/\nu_H}$ as function of $H$, where $\nu_H$ is the correlation-length critical exponent of two-dimensional percolation: for ${H<-1/\nu_\text{2D}^\text{uncorr}}$, ${1/\nu_H=1/\nu_\text{2D}^\text{uncorr}=3/4}$ and for ${-1/\nu_\text{2D}^\text{uncorr} \leq H < 0}$, ${1/\nu_H=-H}$ \cite{Weinrib83, Weinrib84, Schmittbuhl93}.
	We note that these results are similar to measurements in Refs.~\cite{Prakash92, Schmittbuhl93, Roux94, Sandler04, Marinov06}.
	Inset: Number of red sites as function of the lattice size $L$ for the values of $H$ shown in the main plot.
	}
\end{figure}

The backbone becomes more compact with increasing correlation, which is also compatible with the fact that the shortest path fractal dimension is decreasing in this limit (see Fig.~\ref{fig::03_shortest_path_length}).
For the same reason, one would expect the fractal dimension of the set of red sites $d_\text{rs}$ to decrease with increasing $H$.
Coniglio \cite{Coniglio89} has shown that the red site fractal dimension is related to the correlation-length critical exponent $\nu_\text{2D}^\text{uncorr}$ by ${d_\text{rs}=1/\nu_\text{2D}^\text{uncorr}}$.
To test the theoretical predictions in Refs.~\cite{Weinrib83, Weinrib84, Schmittbuhl93} for ${1/\nu_H}$, we measured the red site fractal dimension $d_\text{rs}$ as function of $H$, see Fig.~\ref{fig::red_sites_backbone}.
Although for $H$ approaching zero the finite size effects become more severe, see inset of Fig.~\ref{fig::red_sites_backbone}, the relation seems to be compatible with the data, in agreement with the results in Refs.~\cite{Prakash92, Schmittbuhl93, Sandler04, Abete04, Marinov06}.
This is consistent with the finite-size scaling in the percolation threshold estimation (see Sec.~\ref{sec::percolation_threshold}).
%
\section{\label{sec::conductivity}Cluster conductivity}
At the percolation threshold, the backbone of the largest cluster is a fractal and the conductivity $C$ between its ends has a power-law dependence on the Euclidean distance $r$ of the end sites,
\begin{equation}
        C(r) \sim r^{-t_H/\nu_H},
\end{equation}
where $t_H$ is the conductivity exponent and we call ${t_H/\nu_H}$ the reduced conductivity exponent \cite{Hong84, Derrida82, Zabolitzky84, Normand88, Lobb84, Frank88, Prakash92, Grassberger99, Pose12}.
For uncorrelated percolation, ${t_\text{2D}^\text{uncorr}/\nu_\text{2D}^\text{uncorr}}=0.9826\pm0.0008$ \cite{Grassberger99}.
As the backbone becomes more compact with increasing correlation (see Sec.~\ref{sec::burning_exponents}), one might expect the conductivity to decay more slowly with the spatial separation, and, consequently, that ${t_H/\nu_H}$ decreases \cite{Prakash92, Andrade11b}.

\begin{figure}
        \includegraphics[width=\columnwidth]{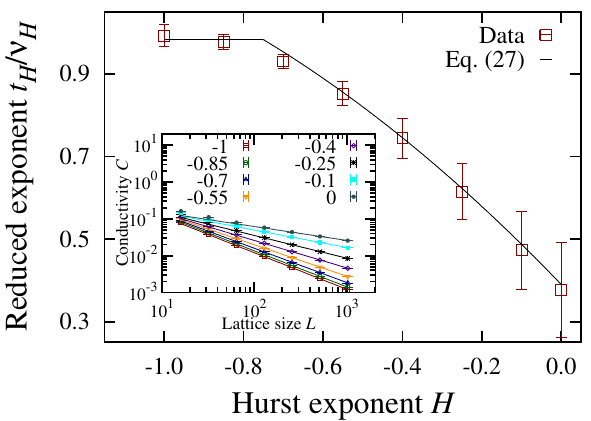}
        \caption{\label{fig::reduced_conductivity_exponent_H}
        (Color online)
        Reduced conductivity exponent ${t_H/\nu_H}$ as function of the Hurst exponent $H$.
        For increasing value of $H$, as the backbone becomes more compact (see Fig.~\ref{fig::06_backbone_size}), ${t_H/\nu_H}$ decreases.
        For uncorrelated disorder, we find $ t_H/{\nu_H}(-1)=-0.992 \pm 0.027$ in agreement with Ref.~\cite{Grassberger99}.
        The solid line corresponds to the expression $t_H/{\nu_H} = {16}/{41}-H-{7H^2}/{25}$ in the range ${-1/\nu_\text{2D}^\text{uncorr} \leq H \leq 0}$ and $t_H/{\nu_H} = t/\nu_\text{2D}^\text{uncorr}$ for $-1 \leq H \leq -1/\nu_\text{2D}^\text{uncorr}$.
        Inset: Conductivity $C$ as function of the lattice size $L$, for the same values of the Hurst exponent $H$ as in the main plot.
        }
\end{figure}

To measure the conductivity $C$ of the backbone, we solved Kirchhoff's laws and obtained for every site $i$ in the backbone:
\begin{equation}
        \sum_k (V_i-V_k) = 0,
\end{equation}
where the sum runs over the nearest neighbors $k$ belonging to the backbone of site $i$, and the conductivity is unity between neighboring sites.
The boundary conditions are chosen such that ${V=N}$ on the top end of the backbone and ${V=0}$ on its bottom end.
Solving the sparse linear system of equations one obtains the conductivity and the value of the potential at each site of the backbone (for details, see e.g. Ref.~\cite{Pose12}).
The inset of Fig.~\ref{fig::reduced_conductivity_exponent_H} shows the conductivity $C$ as function of the lattice size $L$, for different values of $H$.
Since in our setup the distance between the end points ${r \sim L}$, we use this scaling to determine the reduced conductivity exponent ${t_H/\nu_H}$, see Fig.~\ref{fig::reduced_conductivity_exponent_H}.
Our result for uncorrelated percolation agrees with the literature and one observes $t_H/\nu_H$ to decrease with increasing $H$.
We propose the following functional dependence of the reduced conductivity exponent on $H$ (in the range ${-1/\nu_\text{2D}^\text{uncorr} \leq H \leq 0}$):
\begin{equation}
        \label{eqn::conductivity_conjecture}
        \frac{t_H}{\nu_H} = \frac{16}{41}-H-\frac{7H^2}{25}.
\end{equation}
%
\section{\label{sec::ranked_surfaces}Bridge site growth}
\begin{figure}
	\includegraphics[width=\columnwidth]{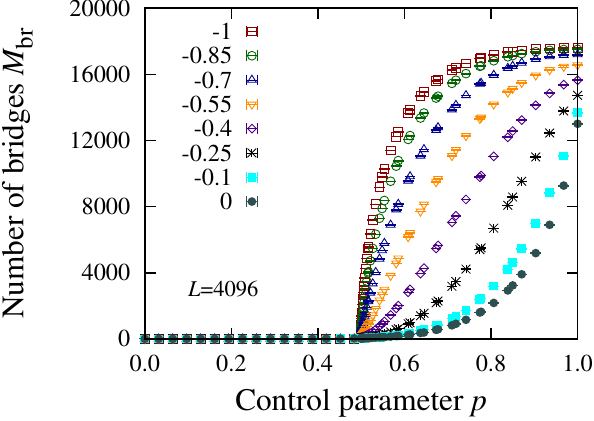}
	\caption{
	\label{fig::coper_bridge_site_growth_with_p}
	(Color online)
	Number of bridge sites $M_\text{br}$ as function of the control parameter $p$, for different values of the Hurst exponent $H$, on a lattice of size ${L=4096}$.
	Results are averages over $10^4$ samples.
	}
\end{figure}

To explore further the impact of correlations on the structure of percolation clusters, we analyze the bridge sites, which are related to red sites, at the percolation threshold \cite{Schrenk12, Fehr11c, Coniglio89}.
Consider the following modification of the percolation model:
While the sites are sequentially occupied, starting from the empty lattice, if a site would lead to the emergence of a spanning cluster between the top and bottom sides of the lattice, this site does not become occupied and is labeled as a bridge site \cite{Schrenk12, Schrenk12b, Schrenk13}.
While the fraction of occupied sites $p$ is lower than the percolation threshold $p_c$, the set of bridge sites is empty, since there would be no percolating cluster in classical percolation for ${p<p_c}$ \cite{Stauffer94}.
At the threshold, the number of bridge sites $M_\text{br}$ behaves identically to the number of red sites and diverges with the lattice size as
\begin{equation}
	M_\text{br} \sim L^{1/\nu},
\end{equation}
where $\nu$ is the correlation-length critical exponent of percolation \cite{Schrenk12, Coniglio89, Scholder09}.
For uncorrelated disorder, at ${p>p_c}$, the number of bridge sites grows as a power law with the distance from the threshold,
\begin{equation}
	\label{eqn::power_law_growth}
	M_\text{br} \sim (p-p_c)^{\zeta},
\end{equation}
where ${\zeta=0.50\pm0.03}$ \cite{Schrenk12} is called the bridge growth exponent (see also Fig.~\ref{fig::coper_bridge_site_growth_with_p}).
When $p$ goes to unity, the set of bridge sites merge to a singly connected line, spanning the lattice horizontally, which is the watershed of the landscape of considered heights $h$, if the top and bottom sides of the lattice would be connected to water outlets \cite{Fehr09, Fehr11, Daryaei12, Andrade13, Daryaei13, Daryaei13b}.
For uncorrelated landscapes, this watershed is a fractal path of dimension ${d_\text{br}=1.2168\pm0.0005}$ \cite{Fehr11c}.

\begin{figure}
	\includegraphics[width=\columnwidth]{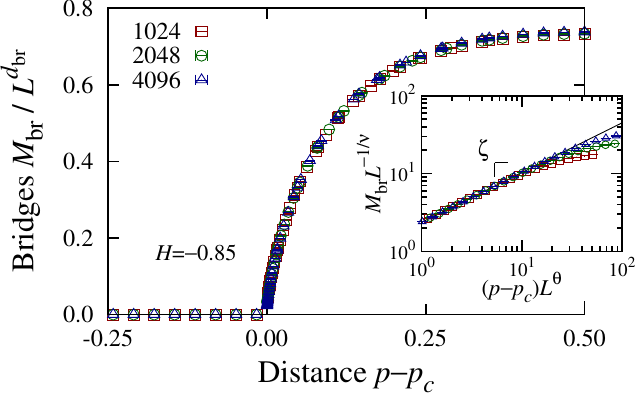}
	\caption{
		\label{fig::coper_bridge_site_growth_H-085}
		(Color online)
		Main plot: Rescaled number of bridge sites $M_\text{br}/L^{{d_\text{br}}}$ as function of the distance to the percolation threshold ${p-p_c}$, with ${H=-0.85}$, for different lattice sizes $L$.
		Here, we use ${d_\text{br}(-0.85)=1.211}$ \cite{Oliveira11, Fehr11b, Fehr11c}.
		Inset: Rescaled number of bridge sites ${M_\text{br}L^{-1/\nu_\text{2D}^\text{uncorr}}}$ as function of the scaling variable ${(p-p_c)L^\theta}$, with ${\theta=0.72}$.
		The solid line is a guide to the eye with slope $0.64$.
	}
\end{figure}

To determine how the bridge site growth depends on $H$, we measured the number of bridge sites $M_\text{br}$ as function of $p$, for different values of $H$, see Fig.~\ref{fig::coper_bridge_site_growth_with_p}.
For values of ${H\leq-1/\nu_\text{2D}^\text{uncorr}}$, we observe that the data for different lattice sizes collapses, when rescaled by $L^{d_\text{br}(H)}$, for all values of ${p>p_c}$ (see Fig.~\ref{fig::coper_bridge_site_growth_H-085}).
This suggests that the same crossover scaling as in the uncorrelated case \cite{Schrenk12} can be applied to extract the growth exponent $\zeta$:
\begin{equation}
	M_\text{br}(p,L) = L^{1/\nu_\text{2D}^\text{uncorr}} F[ (p-p_c)L^\theta ],
\end{equation}
where the scaling function ${F[x] \sim x^\zeta}$ for large $x$ and the power-law behavior of $M_\text{br}$ in the lattice size $L$ and $p$ yields
\begin{equation}
	\label{eqn::exp_scaling_crossover}
	\theta = (d_\text{br}-1/\nu_\text{2D}^\text{uncorr})/\zeta.
\end{equation}
For ${H=-0.85}$, the rescaled data is shown in the inset of Fig.~\ref{fig::coper_bridge_site_growth_H-085} and the growth exponent is ${\zeta(-0.85)=0.64\pm0.06}$, larger than for ${H=-1}$.
The corresponding value of $\theta$ yielding the best collapse of the data is ${\theta=0.72\pm0.08}$, in agreement with the scaling relation of Eq.~(\ref{eqn::exp_scaling_crossover}), given the known dependence of the watershed fractal dimension $d_\text{br}$ on $H$ \cite{Oliveira11, Fehr11b, Fehr11c}.

\begin{figure}
	\includegraphics[width=\columnwidth]{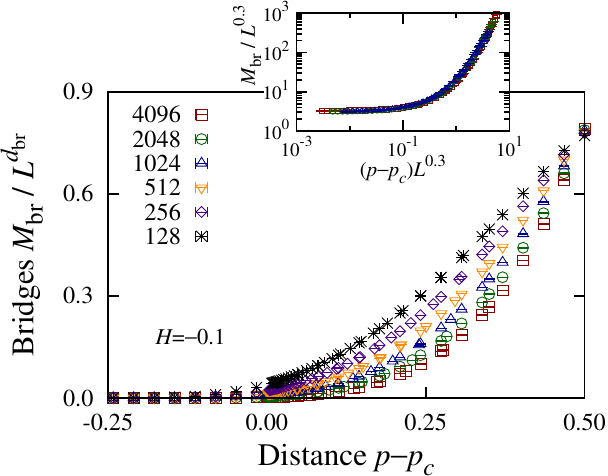}
	\caption{
		\label{fig::coper_bridge_site_growth_H-0.1}
		(Color online)
		Main plot: Rescaled number of bridge sites $M_\text{br}/L^{{d_\text{br}}}$ as function of ${p-p_c}$, with ${H=-0.1}$, for different lattice sizes $L$.
		Inset: Data for the largest three $L$, with $M_\text{br}/L^{0.3}$ as function of $(p-p_c)L^{0.3}$.
	}
\end{figure}

\begin{figure}
	\includegraphics[width=\columnwidth]{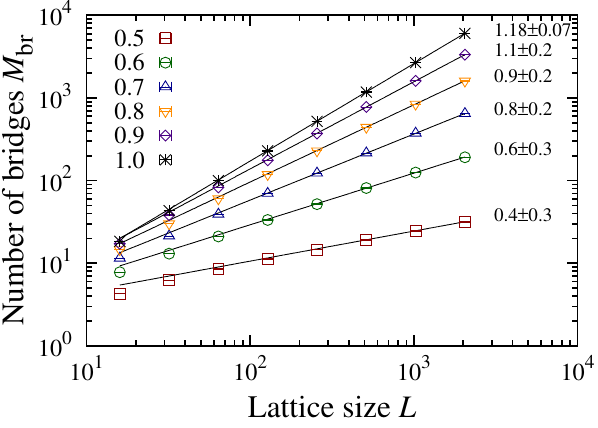}
	\caption{
		\label{fig::bridge_site_growth_finite_size_effects_H-01}
		(Color online)
		Number of bridge sites $M_\text{br}$, for ${H=-0.1}$, as function of the lattice size, for different values of the fraction of occupied sites ${p=p_c=0.5}$, $0.6$, $0.7$, $0.8$, $0.9$, and unity.
		The solid lines are guides to the eye.
		The estimated slopes are indicated on the right had side of the figure.
	}
\end{figure}

For ${H\geq-1/\nu_\text{2D}^\text{uncorr}}$, the behavior of bridge sites is qualitatively different from the uncorrelated case.
The rescaled number of bridge sites ${M_\text{br}(p)/L^{d_\text{br}(H)}}$ does not overlap for different lattice sizes $L$ for any value of ${p>p_c}$, except when the complete fractal line has emerged, i.e. for ${p\to1}$.
An example of this behavior, for ${H=-0.1}$, is shown in Fig.~\ref{fig::coper_bridge_site_growth_H-0.1}.
To analyze this size effect in more detail, we plot in Fig.~\ref{fig::bridge_site_growth_finite_size_effects_H-01} the number of bridges $M_\text{br}$ as function of the lattice size $L$, for different values of ${p \geq p_c}$.
One observes that, in contrast to the uncorrelated case \cite{Schrenk12}, for ${p > p_c}$, there is no crossover to the fractal dimension of the continuous bridge line $d_\text{br}$.
Precisely at the critical point, the expected behavior ${M_\text{br} \sim L^{1/\nu_H}}$ is still observed.
%
\section{\label{sec::fin}Final remarks}
Concluding, we studied percolation with long-range correlation in the site occupation probabilities, as characterized by the Hurst exponent $H$.
The site percolation threshold of the triangular lattice was argued to be ${1/2}$, independent of $H$.
For $H$ approaching zero the fractal dimension of the largest cluster, as well as the exponent ratio ${\gamma_H/\nu_H}$ where found to increase in accordance with the hyperscaling relation.
The fractal dimensions of the complete and the accessible perimeter were observed to approach ${3/2}$ for ${H\to0}$, while the duality relation between both exponents seems to hold independently on the value of $H$.
As $H$ increased, the backbone of the largest cluster was observed to become more compact, consistent with the scaling behavior of shortest path, red sites, and conductivity.
Finally, we found the bridge growth exponent to increase with increasing $H$.
While the qualitative picture is consistent with previous studies in the literature, we proposed quantitative relations for the dependence of the critical exponents of the percolation transition on $H$, as being the simplest rational expressions that fit the numerical data.
\begin{acknowledgments}
We acknowledge financial support from the ETH Risk Center, the Brazilian institute INCT-SC, and grant number FP7-319968 of the European Research Council.
\end{acknowledgments}
\bibliography{bibliography.bib}
\end{document}